\RequirePackage{fix-cm}
\documentclass{svjour3}                     
\smartqed  
\usepackage{graphicx}
\usepackage{natbib}
\begin{document}

\title{Which type of planets do we expect to observe in the Habitable Zone?
}


\author{Vardan Adibekyan$^{1}$	\and
	 Pedro Figueira$^{1}$		\and
	 Nuno C. Santos$^{1,2}$
}


\institute{Vardan Adibekyan \at
            \email{Vardan.Adibekyan@astro.up.pt}           
           \and
              $^{1}$	Instituto de Astrof\'isica e Ci\^encias do Espa\c{c}o, Universidade do Porto, CAUP, Rua das Estrelas, 4150-762 Porto, Portugal \\
              $^{2}$	Departamento de F\'isica e Astronomia, Faculdade de Ci\^encias, Universidade do Porto, Rua do Campo Alegre, 4169-007 Porto, Portugal
}

\date{Received: date / Accepted: date}

\maketitle

\begin{abstract}
We used a sample of super-Earth-like planets detected by the Doppler spectroscopy and transit techniques to explore the dependence of orbital parameters of the planets on
the metallicity of their host stars. We confirm the previous results (although still based on small samples of planets) that  super-Earths orbiting around metal-rich stars are not observed to be as distant from their 
host stars as we observe their metal-poor counterparts to be. The orbits of these super-Earths with metal-rich hosts usually do not reach into the Habitable Zone (HZ), 
keeping them very hot and inhabitable. We found that most of the known planets in the HZ are orbiting their GK-type hosts which are metal-poor. The metal-poor
nature of planets in the HZ suggests a high Mg abundance relative to Si and high Si abundance relative to Fe. These results lead us to speculate that
HZ planets might be more frequent in the ancient Galaxy and had compositions different from that of our Earth.

\end{abstract}

\section{Introduction}
\label{intro}
Twenty years ago, in 1995, the first extrasolar planet orbiting a main sequence solar-type star, 51 Pegasi, was detected \citep{Mayor-95}. 
Nowadays, several thousand of exoplanets and  exoplanet candidates are announced,  dozens of them being located in the so called habitable 
zone: a zone where exoplanets permitted to  have liquid water on their 
solid surface \citep[][]{Cruz-13, Kopparapu-13}. 
This large amount of discovered exoplanets allowed to achieve an unprecedented advancement on our understanding of formation and evolution 
of exoplanets, although,  sometimes, newly discovered planets bring more questions than answers. The recorded progress was possible due 
to many observational studies of individual and statistical properties of exoplanets (and their host stars), successfully  followed by 
theoretical explainations. 

One of the first observed properties of exoplanets was the giant-planet 
occurrence dependence on the host star metallicity \citep[e.g.][]{Gonzalez-97, Santos-01, Santos-04, Johnson-10, Mortier-13}. 
Interestingly, this dependence, if exist, is likely very weak for low-mass/small-size planets \citep[e.g.][]{Sousa-11, Buchhave-15}.
These observational results were theoretically explained within the context of core-accretion theory of planet formation
\citep[][]{Ida-04, Mordasini-12}. It is worth noting that these correlations were recently reproduced in the context of gravitational instability too.
This was done by \citet{Nayakshin-15}, who used a Tidal Downsizing hypothesis for planet formation.

The importance of stellar (and disk) metallicity is probably not only limited to the formation of planets. Recently, it was shown that the architecture of 
planets may also depend on the stellar metallicity \citep[][]{Beauge-13, Dawson-13, Adibekyan-13}.
Moreover, \citet[][]{Dawson-15} proposed that the presence or absence of gaseous atmosphere of small-sized planets depends on 
metallicity thorough disk solid surface density.

In all the aforementioned studies, the iron content was used as a proxy for overall metallicity. However, recent works
showed that elements other than iron may play a very important role for planet formation. In particular, it was shown that 
iron-poor stars hosting giant \citep[][]{Haywood-08, Haywood-09, Adibekyan-12a} and low-mass \citep[][]{Adibekyan-12b} planets 
are systematically enhanced in $\alpha$-elements. It was also shown that low-mass planet hosts show a high Mg/Si abundance ratio 
compared to the field stars without any detected planetary companion \citep[][]{Adibekyan-15}.

The importance of individual heavy elements and specific elemental ratios is not only limited to the 
formation of the planets, but may also control the structure and composition of the planets 
\citep[e.g.][] {Grasset-09, Bond-10, Delgado-10, Rogers-10, Thiabaud-14, Thiabaud-15, Dorn-15, Kereszturi-16}. In particular,  
Mg/Si and  Fe/Si  mineralogical ratios were proposed to allow to constrain the internal structure of terrestrial planets \citep[][]{Dorn-15}. 
These theoretical models were recently successfully tested on three terrestrial planets  by \citet[][]{Santos-15}.

In this work we will first revisit the results of \citet[][]{Beauge-13} and \citet[][]{Adibekyan-13} on the dependence 
of orbital distance ($a$) of small-size/low-mass planets on metallicity. Then,  by extrapolating our results, we  discuss the possible properties of planets in the HZ.

\section{Super-Earths on the period -- mass/radius diagram}
\label{sec:1}

The recent works of \citet[][]{Beauge-13} and \citet[][]{Adibekyan-13} indicate that there is a dependence of orbital distances of super-Earths and Neptune-like planets
on the stellar metallicity. \citet[][]{Beauge-13}, by using a sample of exoplanet candidates from the $Kepler$ mission, found that small planets ($R_{p} < 4R_{\oplus}$)
orbiting around metal-poor stars always have periods greater than 5 days, while planets with similar sizes but around metal-rich stars show shorter periods.
\citet[][]{Adibekyan-13}, by using a sample of radial velocity (RV) detected exoplanets, found that low-mass planets around metal-rich stars do not show large
orbital periods (greater than ~20 days), while planets of the same mass, but orbiting around metal-poor stars span a wider range of orbital periods
(see their Fig. 2).
These two results are not fully compatible and may come from the possible detection biases in the RV or transit surveys, from the way how the samples were built,
or from the low-number statistics. However, they both suggest that the period (or orbital distance) distribution of the observed planets depends on the stellar metallicity.

\begin{figure}
\begin{tabular}{cc}
  \includegraphics[width=0.5\textwidth]{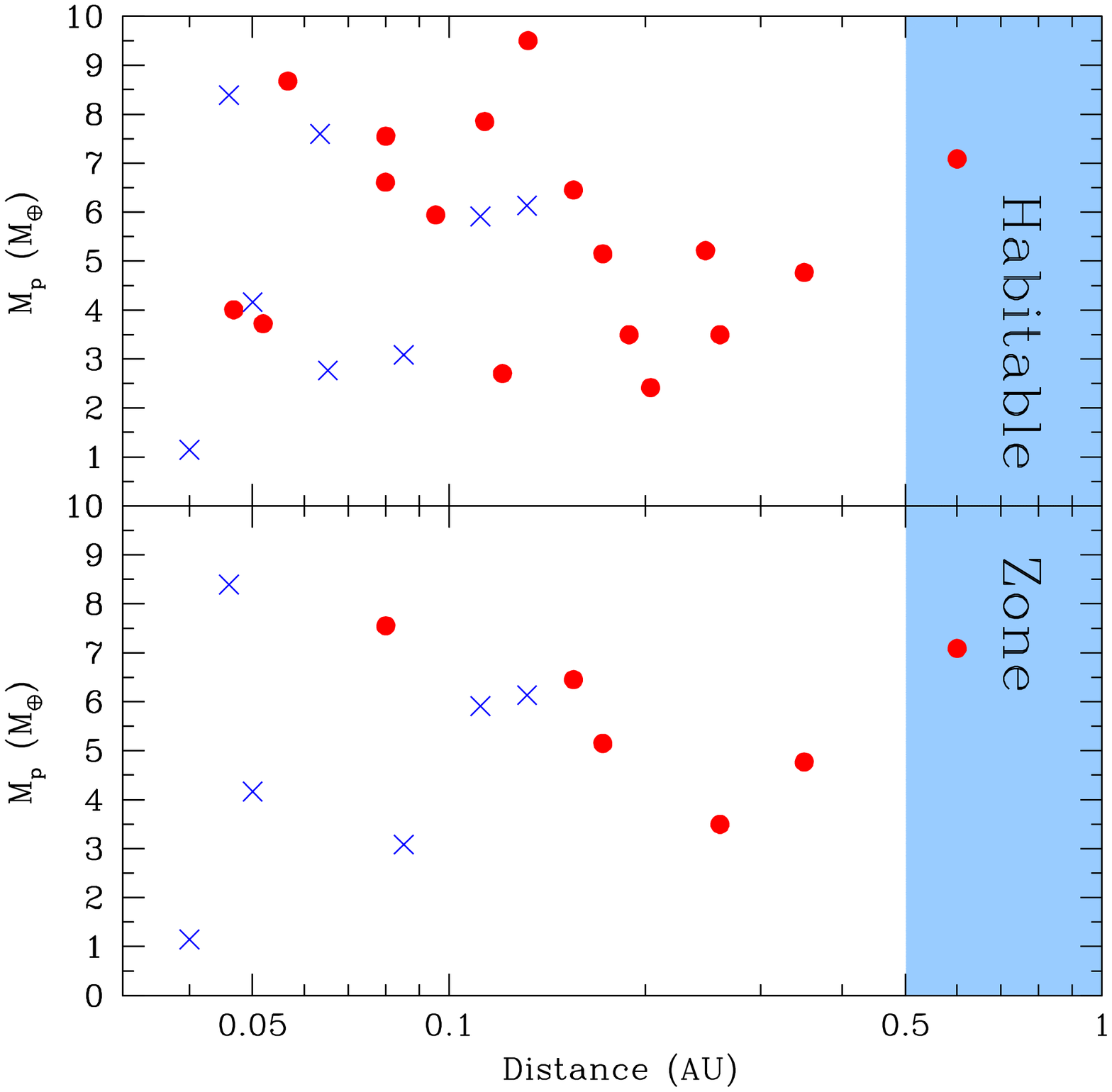}
  \includegraphics[width=0.5\textwidth]{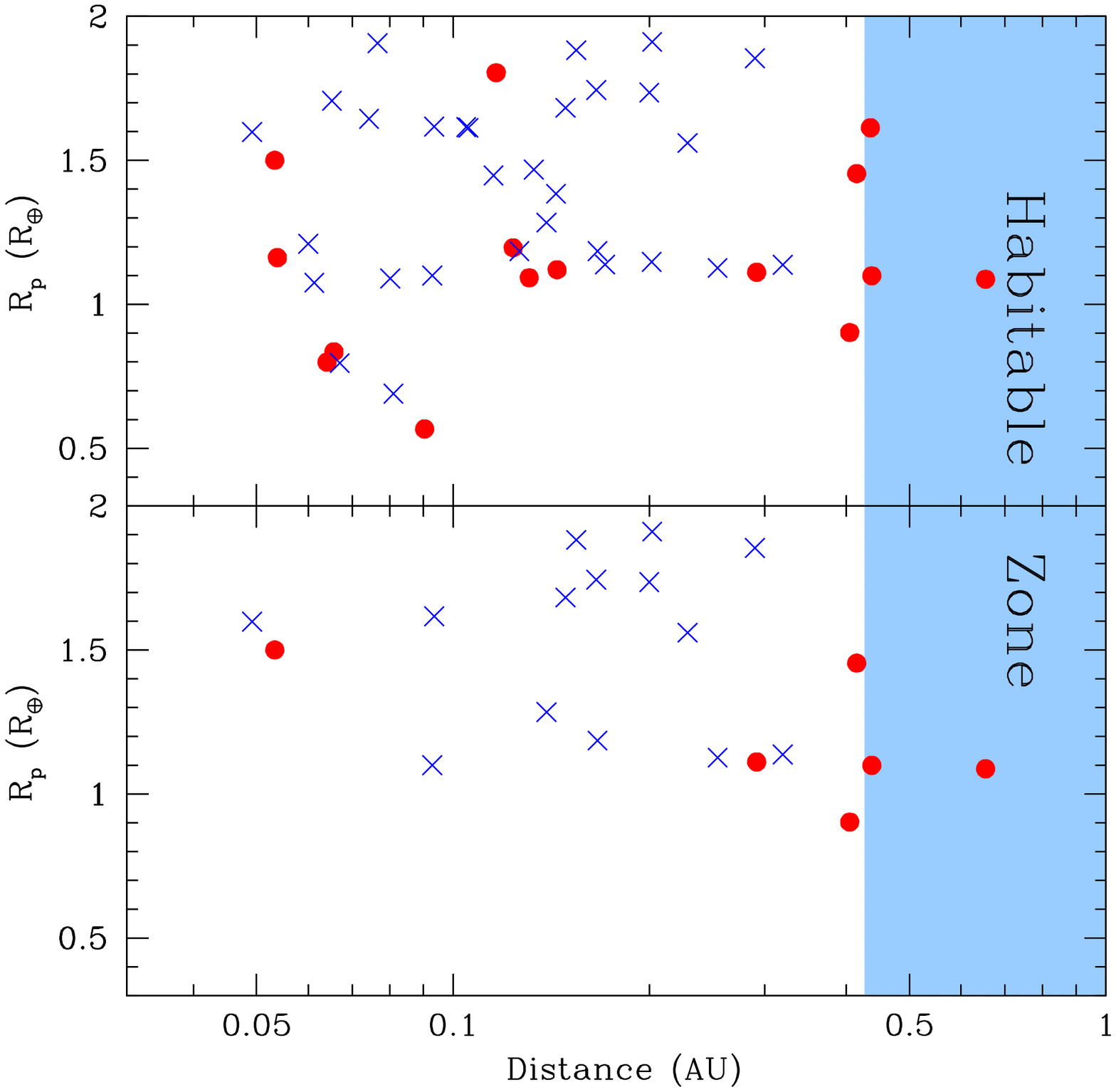}
\end{tabular}
\vspace{-0.5cm}
\caption{The position of low-mass planets around FGK dwarf stars on the $a$ -- Mp plane (left panel) and of 
small-size planets around FGK dwarf stars on the $a$ -- Rp plane (right panel). The bottom panels show the position of the planets
at the largest orbital distances in the systems. Red circles correspond to planets orbiting stars with [Fe/H] $\leq$ -0.1 dex and 
blue crosses represent planets orbiting metal-rich stars with [Fe/H] $>$ -0.1 dex. 
The HZ of the star with the shortest inner edge distance from the star is presented in blue shade.}
\label{fig:1}       
\end{figure}

\subsection{Distance -- mass diagram for low-mass planets}
\label{subsec:1.1}

To revisit the results of \citet[][]{Adibekyan-13}, we selected all the RV detected low-mass planets (exoplanet.eu\footnote{http://exoplanet.eu/}) 
around FGK dwarf stars ($M_{*} > 0.5M_{\odot}$) for which
stellar parameters were derived in a homogeneous way (SWEET-Cat: \citet{Santos-13})\footnote{
https://www.astro.up.pt/resources/sweet-cat}. Since, low-mass planets around metal-rich stars may have higher mass  planetary companions at larger distances that
may affect (due to gravitational interactions) the orbital distances of the planets \citep[][]{Adibekyan-13}, 
we selected only those systems where the highest mass planetary companion has $M_{p} \leq 10M_{\oplus}$.

Our sample consists of 26 planets in 12 systems. On the top-left panel of Fig.~\ref{fig:1} we show the position of all the planets on the 
$a$ -- $M_{p}$ diagram, separating planets by their host star's metallicity. The bottom-left panel shows positions of the planets with the largest orbital 
distances in the system. The figure clearly hints a lack of planets at large orbital distances from their metal-rich hosts and confirms the 
result of \citet[][]{Adibekyan-13}. 

To evaluate the statistical significance of this result we applied a simple binomial statistics test.
All the RV detected planets orbiting around metal-rich stars (9 planets)
are within ~0.13 AU distance from their hosts (see the top-left panel of Fig. 1). There are also 9 planets orbiting around metal-poor stars within the same distance.
Thus, if we assume that the metallicity is not the parameter that determines the positions of these planets in the plot, then the probability of a planet to orbit
around a star within the mentioned distance is (9+9)/26 (26 is the total number of planets). Under this assumption our data follows the binomial distribution and
the binomial statistics give a probability of 
P$_{bin}$=0.039 that all the planets orbiting metal-rich stars would orbit within ~0.13 AU distance from their hosts by chance. 
The same statistical test when applied to the data from the bottom-left panel of Fig. 1 provide probability of P$_{bin}$=0.036.

A quite common test that could be performed to our data is the bootstrapping. If we shuffle the whole data large number of times and count the number of trials that all the planets 
around metal-rich stars have orbits closer than ~0.13 AU we can calculate the probability that the event occurred by chance \citep[see e.g.][]{Adibekyan-13}. However,
these probability values will be very close to the p-values obtained from the binomial statistics, because the underlying assumption is the same.

We would like to stress that the samples are small, therefore, the results and conclusions regarding them should be considered with caution.

It is also very interesting to note that planets around metal-rich stars are usually in single or double systems, while metal-poor stars
host more planets (up to six planets -- HD40307: \citep{Tuomi-13}). 

\subsection{Distance -- radius diagram for small-size planets}
\label{subsec:1.2}

We constructed our transit sample from that of \citet{Buchhave-14}, by selecting planetary systems that contains only small-size planets 
of $R_{p} \leq 2R_{\oplus}$ and orbiting around FGK dwarfs. We note that this radius is an approximate maximum radius for habitable planets as suggested by 
\citet{Alibert-14}.
To decrease the possible false-positive rates in our sample, we selected only planets which are confirmed 
(NASA Exoplanet Archive\footnote{NASA Exoplanet Archive -- http://exoplanetarchive.ipac.caltech.edu/}) 
or planets which are in multiple systems \citep[see][]{Lissauer-12}. Following \citet[][]{Buchhave-14}, we excluded systems 
with highly irradiated planets (stellar flux F $>$ 5$\times10^{5}$ J s$^{-1}$ m$^{-2}$), that might have undergone significant atmospheric evaporation 
\citep[see also][]{Owen-13}. Our final sample consists of 45 planets in 20 systems.

The distributions of all the planets and the planets with the largest orbital distance in a system on the $a$ -- R$_{p}$ plane are
shown on the top-right and bottom-right panels of Fig.~\ref{fig:1}. As for the RV detected planets, the planets are separated
according to their host star metallicity. The distribution of the $a$ of the transiting planets is qualitatively similar to that of RV detected planets.
Planets around metal-rich stars do not orbit their stars at distances as large as their metal-poor counterparts. However, orbital distances of the 
metal-rich transiting planets are slightly larger than that observed for metal-rich planets detected with the Doppler spectroscopy.

We applied the same binomial statistics as it was done for the RV detected planets and obtained P$_{bin}$ = 0.029 and  P$_{bin}$ = 0.044 for the samples of
the top-right and bottom-right panels of Fig. 1, respectively. We again would like to remind the reader about the limited size of the samples.

It is interesting to see that the fraction of metal-poor planets is much lower in the transiting data than in the RV sample. This might be related to the different 
fields (e.g. different stellar populations) surveyed by the RV and transit search programs, and/or shift of the zero-points of the metallicities. However, 
\citet{Adibekyan-12a} showed that there is a good agreement between metallicities derived by \citet{Buchhave-12} and by standard spectroscopic techniques based
on curve of growth \citep[e.g.][]{Sousa-08}. It is also still possible that some of the transiting planet candidates are false positive that affects the 
true metallicity distribution of the transiting samples.

We note that \citet[][]{Beauge-13} did not find small planets around metal-poor stars with short orbital periods, likely because their sample was
about one third of this sample. 

\begin{figure*}
  \includegraphics[angle=270, width=1.0\textwidth]{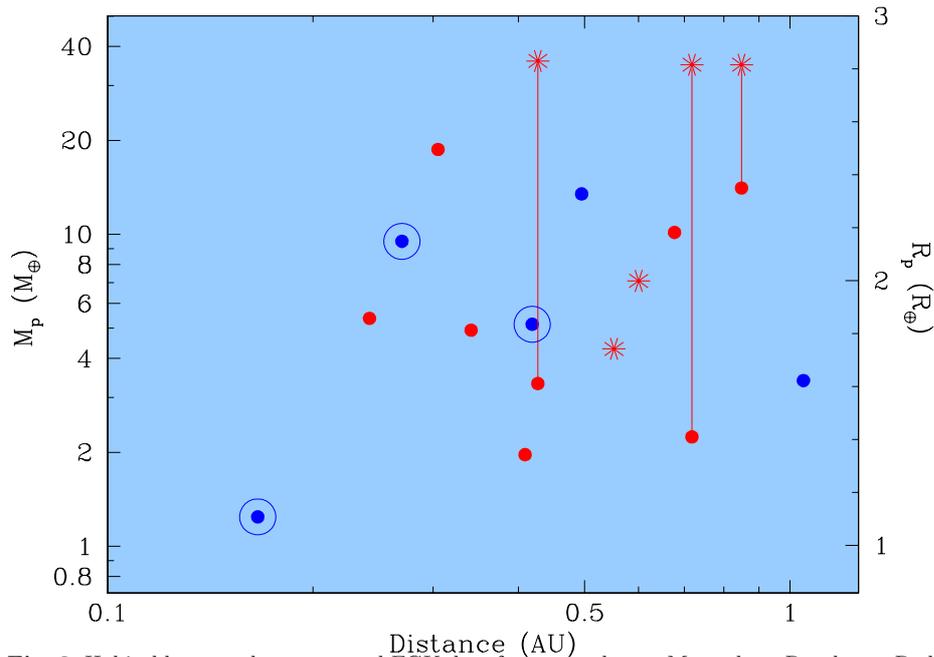}
\vspace{-0.5cm}
\caption{Habitable zone planets around FGK dwarf stars on the $a$ -- M$_{p}$ and $a$ -- R$_{p}$ planes. 
Red symbols correspond to planets orbiting stars with [Fe/H] $\leq$ -0.1 dex and 
blue symbols represent planets orbiting metal-rich stars with [Fe/H] $>$ -0.1 dex.
Planets with mass measurements are presented by asterisks and with radius measurements by filled circles. The three planets with both mass and radius 
measurements are connected by red lines. Three planets orbiting the coolest stars are marked by open large circles.}
\label{fig:2}       
\end{figure*}

\subsection{Planets in the habitable zone}
\label{subsec:1.3}

In Fig.~\ref{fig:1}, we show the HZ of the star with the shortest inner edge distance from the star. The HZ of the RV detected planet hosts are calculated 
following \citet{Kopparapu-13}. For the transiting planets it is hard to get precise information on the host stars luminosity, because of their poor distance estimation. 
In the right panel of the plot the HZ starts from $a$=0.427 AU, which is the orbital distance of Kepler-62e, a habitable planet around K type star \citep{Borucki-13}. 
As one can see, most of the planets orbit their stars in the zone where the temperature is still high for water to be in a liquid form at the surface of the planets.
While the lack of long-distance super-Earths is mostly due to the detection limits in the planet search programs, this bias probably is not responsible 
for the sub-grouping of planets depending on their host stars metallicity. Fig.~\ref{fig:1} shows that metal-poor super-Earths are closer to 
the HZ than their metal-rich counterparts, and the two planets (Kepler-62e and HD40307g), that are located in the HZ, are hosted by metal-poor stars.

In Fig.~\ref{fig:2}, we plot all the known planets in the HZ that are orbiting FGK type stars\footnote{http://phl.upr.edu/projects/habitable-exoplanets-catalog}
on the $a$ -- M$_{p}$ and $a$ -- R$_{p}$ planes. Ten out of the 15 planets (note, that three planets are plotted twice because they have 
both radius and mass measurements) are orbiting stars with [Fe/H] $<$ -0.1 dex. In fact, all the five planets for which there is a RV confirmation are orbiting 
low-metallicity stars. Three of the planets
orbiting metal-rich stars, have the coolest hosts in the sample with the T$_{eff} <$  4050 K, for which the derivation of stellar parameters, including metallicity
is more difficult, hence probably less precise. 

\section{Concluding remarks and outlook}
\label{sec:2}

Our samples of low-mass and small-size planets detected by the RV and transit methods show that planetary architecture depends on the metallicity of the disk
where they formed. In particular, we showed that these super-Earths orbiting around metal-rich stars ([Fe/H] $>$ -0.1 dex) do not have orbits as large as 
it is observed for their metal-poor counterparts. The orbits of the super-Earths with metal-rich hosts, 
usually do not reach to the HZ, making them very hot and inhabitable. We note that the maximum mass and radius of the selected planets are 10M$_{\oplus}$ and
2R$_{\oplus}$, which are approximate limits for habitable planets \citep{Alibert-14}. 

The sample of 15 planets orbiting their FGK hosts inside the HZ (all the known planets), shows that these planets tend to orbit stars with low metallicities: 
only three out of the 15 planets have hosts with metallicity higher than that of the Sun. The only planet in the HZ orbiting solar-like (G-type) 
metal-rich star is the very recently discovered Kepler-452b \citep[][]{Jenkins-15}.

The extrapolation of our results, that the planets in the HZ tend to orbit around metal-poor stars,  can have very interesting and important implications.
As it was shown in \citet{Adibekyan-12a}, metal-poor hosts of super-Earths tend to be enhanced in $\alpha$-elements, which means high Si/Fe ratio. 
Similarly, metal-poor low-mass planet hosts are more enhanced in Mg relative to Si, i.e, high Mg/Si ratio \citep{Adibekyan-15}. These two mineralogical
ratios, Si/Fe and Mg/Si, are very important for the formation of terrestrial planets \citep[e.g.][]{Bond-10, Dorn-15}. Moreover, the structure and composition of 
the planets is controlled by these ratios \citep[][]{Dorn-15}. 

The discussion presented above, on the dependence of composition of planets on the chemical properties of their hosts, leads us to speculate that probably
the frequency of planets in the HZ was higher in the ancient Galaxy and in the outer disk of the Galaxy, when/where the metallicity is on average lower than in the 
solar neighborhood. 
Moreover,  most of these planets in the HZ (because of lower metallicity, high Si/Fe, and high Mg/Si) should have composition 
that might be very different than that of our Earth. 

Our results and discussion is based on a small sample of low-mass/small-size planets and some of the conclusions are, of course, very speculative. We have an
example of our Earth --  a habitable planet in the HZ -- which is orbiting a non-metal-poor star. The future large planet search missions,
such as TESS, CHEOPS, and PLATO-2.0 will certainly help us to make the picture more clear and to understand if the case of the Earth is a rule or rather an exception.

\begin{acknowledgements}
This work was supported by Funda\c{c}\~ao para a Ci\^encia e a Tecnologia (FCT) through the research grant UID/FIS/04434/2013.
V.A. acknowledges the support from the Funda\c{c}\~ao para a Ci\^encia e Tecnologia, FCT (Portugal) in the form of the grant
SFRH/BPD/70574/2010.
P.F. and N.C.S. also acknowledge the support from FCT through Investigador FCT contracts of reference IF/01037/2013 and IF/00169/2012,
respectively, and POPH/FSE (EC) by FEDER funding through the program ``Programa Operacional de Factores de Competitividade - COMPETE''. 
PF further acknowledges support from Funda\c{c}\~ao para a Ci\^encia e a Tecnologia (FCT) in the form of an exploratory project of reference IF/01037/2013CP1191/CT0001. 
This work results within the collaboration of the COST Action TD 1308.
We would like to thank Jo\~{a}o Faria for his interesting comments and suggestions.
\end{acknowledgements}

\bibliographystyle{spbasic}      
\bibliography{adibekyan_bibliography}   


\end{document}